\documentclass[iop,apj,twocolumn,onecolappendix,numberedappendix]{emulateapj}
\usepackage{latexsym,graphicx,rotating,amsmath, epsfig, natbib,hyperref}
\usepackage[usenames, dvipsnames]{xcolor}

\newcommand\Beq{\begin{eqnarray}} 
\newcommand\Eeq{\end{eqnarray}}

\newcommand{\B}{B19}

\newcommand{\Msun}{\ensuremath{{M}_\odot}}

\newcommand{\Lsun}{\ensuremath{{L}_{\odot}}}

\newcommand{\note}[1]{{\color{black} #1}}

\begin{document}

\title{Low-frequency variability in massive stars: \\ Core generation or surface phenomenon?}
\shorttitle{Low-frequency variability: Core or surface?}
\shortauthors{Lecoanet et al.}
\author{Daniel Lecoanet\altaffilmark{1,2}, Matteo Cantiello\altaffilmark{3,2}, Eliot Quataert\altaffilmark{4}, Louis-Alexandre Couston\altaffilmark{5,6,7}, Keaton J.~Burns\altaffilmark{3,8}, Benjamin J.~S.~Pope\altaffilmark{9,10,11}, Adam S.~Jermyn\altaffilmark{3}, Benjamin Favier\altaffilmark{5}, \& Michael Le Bars\altaffilmark{5}}
\altaffiltext{1}{Princeton Center for Theoretical Science, Princeton University, Princeton NJ 08544, USA}
\altaffiltext{2}{Department of Astrophysical Sciences, Princeton Univesity, Princeton NJ 08544, USA}
\altaffiltext{3}{Center for Computational Astrophysics, Flatiron Institute, New York, NY 10010, USA}
\altaffiltext{4}{Astronomy Department and Theoretical Astrophysics Center, University of California, Berkeley, CA 94720, USA}
\altaffiltext{5}{CNRS, Aix Marseille Universit\'e, Centrale Marseille, IRPHE UMR 7342, Marseille 13013, France}
\altaffiltext{6}{British Antarctic Survey, Cambridge CB3 0ET, UK}
\altaffiltext{7}{Department of Applied Mathematics and Theoretical Physics, University of Cambridge, Cambridge CB3 0WA, UK}
\altaffiltext{8}{Department of Mathematics, Massachusetts Institute of Technology, Cambridge, Massachusetts 02139, USA}
\altaffiltext{9}{Center for Cosmology and Particle Physics, Department of Physics, New York University, New York, New York 10003, USA}
\altaffiltext{10}{Center for Data Science, New York University, New York, New York 10011, USA}
\altaffiltext{11}{Sagan Fellow}

\email{lecoanet@princeton.edu}

\begin{abstract}
\citet{Bowman2019} reported low-frequency photometric variability in 164 O- and B-type stars observed with K2 and TESS. They interpret these motions as internal gravity waves, which could be excited stochastically by convection in the cores of these stars. The detection of internal gravity waves in massive stars would help distinguish between massive stars with convective or radiative cores, determine core size, and would provide important constraints on massive star structure and evolution. In this work, we study the observational signature of internal gravity waves generated by core convection. We calculate the \textit{wave transfer function}, which links the internal gravity wave amplitude at the base of the radiative zone to the surface luminosity variation. This transfer function varies by many orders of magnitude for frequencies $\lesssim 1  \, {\rm d}^{-1}$, and has regularly-spaced peaks near $1 \, {\rm d}^{-1}$ due to standing modes. This is inconsistent with the observed spectra which have smooth ``red noise'' profiles, without the predicted regularly-spaced peaks. The wave transfer function is only meaningful if the waves stay predominatly linear. We next show that this is the case: low frequency traveling waves do not break unless their luminosity exceeds the radiative luminosity of the star; and, the observed luminosity fluctuations at high frequencies are so small that standing modes would be stable to nonlinear instability. These simple calculations suggest that the observed low-frequency photometric variability in massive stars is not due to internal gravity waves generated in the core of these stars. We finish with a discussion of (sub)surface convection, which produces low-frequency variability in low-mass stars, very similar to that observed in \citet{Bowman2019} in higher mass stars.
\end{abstract}

\section{Introduction}

Massive stars play an important role in astrophysics, driving galactic turbulence which regulates star formation \citep{Hayward2016}, and enriching the chemical composition of the interstellar medium \citep{Nomoto2013}. It is thus important to test and calibrate models of massive star structure and evolution. Asteroseismology is an unparalleled technique to probe the interiors of stars \citep{Aerts2010}, and space-based photometry has revolutionized our understanding of stellar interiors \citep[e.g.,][]{Aerts2019}.

The detection of internal gravity waves, which propagate to the bottom of a star's radiative zone, provides precious information about the cores and interiors of massive stars \citep[e.g.,][]{Zwintz2017,Ouazzani2019}.
\citet[][hereafter B19]{Bowman2019} detected low-frequency variability in a large population of massive stars \citep[following similar detections in fewer stars by, e.g.,][]{Blomme2011}. In over one hundred massive stars, \B\, detect luminosity variations with roughly constant amplitude for frequencies $\lesssim 1 \, {\rm d}^{-1}$, and power-laws with slope $-2$ to $-3$ at higher frequencies. This ``red noise'' signal is very different from previous detections of discrete peaks in the power spectrum, likely due to linearly unstable g-modes \citep{Papics2017}. \B\ interpret this low-frequency variability as internal gravity waves, which could be excited by core convection.

Much of the early work on wave generation by convection was motivated by the Sun and stars with convective envelopes \citep[e.g.,][]{Press1981,Goldreich1990}. These theories were extended to massive stars with convective cores \citep[e.g.,][]{Lecoanet2013}, and were validated by high-resolution, turbulent, 3D numerical simulations in Cartesian geometry \citep{Couston2018}. They find an excitation spectrum that decreases rapidly for frequencies above the convective turnover frequency. In contrast, work by other authors find extremely shallow wave spectra at high frequencies \citep{Rogers2013,Edelmann2019}. There is currently no consensus on the spectrum of internal gravity waves at the radiative-convective boundary.

Simulations of wave excitation by convection do not follow the propagation of internal waves to the surface. To predict surface luminosity fluctuations, \citet{Samadi2010} and \citet{Shiode2013} solved the wave propagation problem using eigenmodes and the WKB approximation, but only considered standing modes. \note{These calculation cannot predict the wave amplitude at frequencies between or below the frequencies of the standing modes. Thus, they do not replicate} the ``red noise'' signal observed in \B. Here we test whether or not the signal can be explained by \textit{traveling} waves; there is no previous work studying the surface manifestation of traveling waves in massive stars.

\section{Internal gravity wave propagation}\label{sec:propagation}

The luminosity variations at the surface of a star $\delta L(f;R_\star)$ are related to the radial velocity variations at the radiative-convective boundary $\delta u_r(f;r_{\rm RCB})$ by
\Beq\label{eqn:transfer function}
\delta L(f;R_\star) = T(f) \delta u_r(f;r_{\rm RCB})
\Eeq
for linear waves. Here $f$ is the frequency of the wave, and $T(f)$ is the \textit{wave transfer function}. We check the linearity assumption in Section~\ref{sec:breaking}. Equation~\ref{eqn:transfer function} separates out the nonlinear physics related to convection, which determines $\delta u_r(f;r_{\rm RCB})$ and is uncertain, from the linear physics of wave propagation, which gives $T(f)$.

We will discuss the wave transfer function for a $10M_\odot$, solar-metallicity star near the ZAMS. The stellar structure model is calculated using MESA \citep{Paxton2011,Paxton2013,Paxton2015,Paxton2018,Paxton2019}. The supplementary information shows the wave transfer function for stars from $3M_\odot$ to $20M_\odot$, and from the ZAMS to the TAMS. The broad features of the transfer functions are similar in this range of masses and ages.

We calculate the wave transfer function in two ways. First, we use non-adiabatic eigenmodes calculated with GYRE \citep{Townsend2018}. We assume the non-adiabatic eigenmodes form a complete basis, and expand the perturbations in this basis. We force the horizontal displacement with $\exp(-i2\pi f t)$ at a forcing radius $r_f$ above the radiative-convective boundary. This produces a luminosity variation at the surface $L(R_\star)$ with frequency $f$. We then calculate $L(R_\star)$ for multiple forcing radii in a $0.02R_\star$ range above the radiative-convective boundary. We force at multiple radii to avoid the nodes of the eigenfunctions. $T(f)$ is the average of the amplitude of $L(R_\star)$ over these forcing radii.

To validate this calculation, we also time-evolve the linearized non-adiabatic oscillation equations in the Cowling approximation using Dedalus \citep{Burns2019}. We solve the equations from the radiative-convective boundary ($r_{\rm RCB}\approx 0.22 R_\star$) to the bottom of the surface convection zone ($r_{\rm top}\approx 0.98 R_\star$). We cannot include the core or surface convection zones in the simulation; otherwise linearly unstable convective modes dominate. We force the waves with a boundary condition $\delta u_r(r_{\rm RCB})=\sin(2\pi f t)$. After an initial transient, the luminosity at the top of the domain, $\delta L(r_{\rm top})$ also varies sinusoidally, which allows us to measure $T(f)$. We find good agreement between these two methods of calculating the wave transfer function.  The supplementary information includes more details of the calculations, as well as the comparison between them.

\begin{figure}
  \centering
  \includegraphics[width=\linewidth]{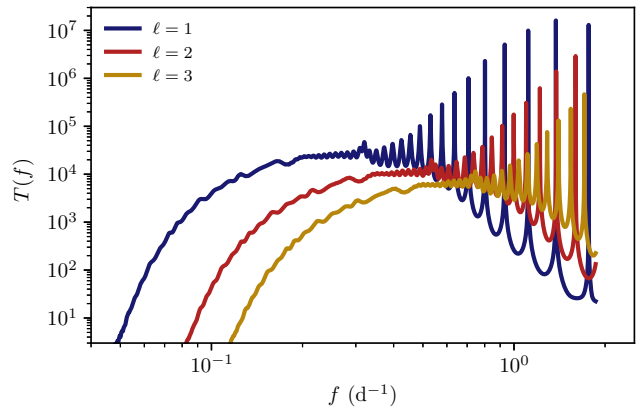}
  \caption{Wave transfer function $T(f)$ for a $10M_\odot$ star near the ZAMS. The transfer function has units of $L/L_\star$ per $u_r(r_{\rm RCB})/(R_\star/{\rm d})$. The transfer function is very small at low frequencies due to radiative damping. At higher frequencies near $1 \, {\rm d}^{-1}$, the transfer function is dominated by sharp peaks associated with standing g-modes.}
  \label{fig:transfer}
\end{figure}

\begin{figure}
  \centering
  \includegraphics[width=\linewidth]{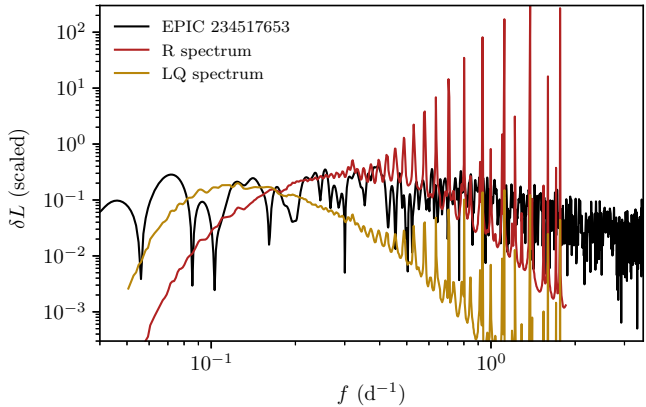}
  \caption{Luminosity variation spectrum for EPIC 234517653 from \B, and two theoretical spectra with different assumptions for the convective excitation spectrum, $\delta u_r(f;r_{\rm RCB})$. Because the theoretical spectra are for linear waves, their overall amplitude is arbitrary. The theoretical spectra show a sharp decline in wave amplitude for waves with $f\lesssim 0.1\,{\rm d}^{-1}$. They also show regularly-spaced peaks at higher frequencies, $f\gtrsim 1\,{\rm d}^{-1}$. Neither of these features are in the observed spectrum.}
  \label{fig:spectra}
\end{figure}

Figure~\ref{fig:transfer} shows the wave transfer function calculated using GYRE for spherical harmonic degrees $\ell=1$, $2$, and $3$. The $\ell=1$ mode dominates the surface luminosity fluctuations (Figure~\ref{fig:spectra}). For each $\ell$, we see the transfer function is smooth at low frequencies, as is observed. However, the transfer function is also very small due to strong wave damping at low frequencies. If the observed low-frequency variability is due to waves from the core, there should be very little power for $f \lesssim 0.1 \, {\rm d}^{-1}$, because these waves are strongly damped. The observed luminosity fluctuations are roughly constant even for $f \lesssim 0.1 \, {\rm d}^{-1}$ (Figure~\ref{fig:spectra} \& \citealt{Blomme2011}). However, it is likely the low-frequency observations are dominated by systematic errors, so this inconsistency at low frequencies does not rule out the detection of propagating internal gravity waves at higher frequencies.

Figure~\ref{fig:transfer} also shows large amplitude, regularly-spaced peaks at frequencies $\gtrsim 1\,{\rm d}^{-1}$. The peaks are at the frequencies of standing g-modes. These frequencies are the resonant frequencies for propagating internal gravity waves. Peaks with this type of regular spacing are not present in the observed data. Thus, we believe the observed low-frequency variability in high-mass stars is not due to linearly propagating internal gravity waves generated in the core.

Because the observed power spectra do not have regularly-spaced peaks, one may be tempted to attribute the low-frequency variability to propagating, rather than standing, internal gravity waves. However, propagating internal gravity waves only produce a smooth spectrum when the damping rate is similar or greater than the mode frequency spacing. This is only true for waves which are strongly damped, which are too low amplitude to be observed. This dichotomy is illustrated by the wave transfer functions in Figure~\ref{fig:transfer}: the transfer function is smooth for low frequencies, but is also very small; whereas at higher frequencies it is dominated by large-amplitude peaks.

To illustrate this point, in Figure~\ref{fig:spectra} we compare the observed spectrum of EPIC 234517653 from B19 to two theoretical spectra. We choose this star because it has spectral type B2, roughly corresponding to our $10M_\odot$ stellar model, and has a relatively simple spectrum without strong rotation effects or signature of linearly unstable modes. Although we plot the ``residual'' spectrum, the ``original'' spectrum is qualitatively similar.

The theoretical spectrum depends on the wave excitation spectrum $\delta u_r(f;r_{\rm RCB})$ via equation~\ref{eqn:transfer function}. To show a range of possibilities, we consider two very different excitation spectra reported in numerical simulations. \citet{Rogers2013} ran a series of 2D cylindrical simulations using the anelastic equations, and found $\delta u_r\sim f^{0.4} \Lambda^{-0.9}$, where $\Lambda=\sqrt{\ell(\ell+1)}$ \citep{Ratnasingam2019}; we refer to results using this spectrum with R. \citet{Couston2018} ran a series of 3D cartesian simulations using the Boussinesq equations, and found $\delta u_r\sim f^{-3.25}\Lambda^{5/2}$; we refer to results using this spectrum with LQ, as they match with the theoretical prediction in \citet{Lecoanet2013}. We then sum luminosity variations over $\ell$ using the \citet{Dziembowski1977} relations as in \citet{Shiode2013}. To maintain the same frequency resolution as the observations, we calculate the maximum brightness fluctuation in frequency bins with the same width as the frequency spacing of the observational data from \B.

Figure~\ref{fig:spectra} shows the luminosity variation spectrum for EPIC 234517653, as well as the R and LQ spectra. \note{The overall normalization of the theoretical spectra are scaled to more easily compare to the observations. The predicted normalization of the LQ spectrum is similar what we plot, whereas the predicted normalization of the R spectrum is about 4 orders of magnitude higher than what we plot (see supplementary information)}. The theoretical spectra inherit two important properties from the wave transfer function (Figure~\ref{fig:transfer}): 1.~At low frequencies ($f\lesssim 0.1 \, {\rm d}^{-1}$), the luminosity perturbations become very small due to strong wave damping; and 2.~At frequencies $\gtrsim 0.5 \, {\rm d}^{-1}$, there are regularly-spaced peaks due to the near-resonant effects of standing modes. Neither of these features are in the observations, which show a fairly flat spectrum at low frequencies, without regularly-spaced peaks at higher frequencies.
\note{Our calculations do not take into account rotation. Rotation would increase the number of resonant peaks at high frequencies, but also decrease the luminosity variation at low frequencies. The supplementary information shows that the high-frequency waves would still manifest as a series of resonant peaks, so rotation does not change the main features of Figure~\ref{fig:spectra}.}

Although here we only compare to a single star, none of the spectra in \B\ show a large number of regularly-spaced peaks similar to the theoretical spectra in Figure~\ref{fig:spectra}. Even stellar power spectra which do have peaks (e.g., EPIC 202061164) only have a couple, and thus do not match the theoretical spectra for convectively-excited internal gravity waves. The standing waves in the R spectrum reach very large amplitudes, which would probably lead to nonlinear interactions. However, in the next section we show that the \textit{observed} variability is too small to be nonlinear.

\section{Internal gravity wave nonlinearity}\label{sec:breaking}

In the previous section, we showed that linearly propagating waves have a surface luminosity spectrum inconsistent with observations. Here we will investigate possible nonlinear effects. We argue that traveling waves do not become strongly nonlinear, i.e., break. While standing waves could in principle experience weak nonlinearities, we show that the observed luminosity fluctuations are so low that an equivalent standing wave would be stable to nonlinear instabilities. Thus, all the peaks of the predicted spectrum should be observable.

\subsection{Traveling wave breaking}

In the absence of damping, waves conserve the \textit{wave luminosity} as they propagate outward,
\Beq\label{eqn:Lwave}
L_{\rm wave} = \frac{4\pi r^2 \rho \omega^4}{Nk_h} |\xi_h|^2.
\Eeq
Here $\xi_h$ is the horizontal wave displacement, $\omega$ is the wave's angular frequency, $N$ is the Brunt-V\"{a}is\"{a}l\"{a} frequency, and $k_h\equiv \Lambda/r$ and $k_r$ are the horizontal and radial components of the wavenumber. In deriving equation~\ref{eqn:Lwave} we have approximated the dispersion relation as $\omega=Nk_h/k_r$.

As waves propagate outward, $\rho$ decreases, leading $\xi_h$ to increase. When $\xi_h k_h\sim 1$, waves break quickly in about a wave period \citep[e.g.,][]{Staquet2002,Liu2010,Eberly2014}. Thus, the criterion for wave breaking is
\Beq\label{eqn:damped}
L_{\rm wave} \gtrsim \frac{4\pi r^2 \rho \omega^4}{Nk_h^3}.
\Eeq
The total wave luminosity is expected to be smaller than the convective luminosity by a factor of the convective Mach number \citep[e.g.,][]{Goldreich1990,Lecoanet2013}; this scaling was verified in the simulations of \citet{Rogers2013} and \citet{Couston2018}.

Radiative damping is significant for some waves. Waves are weakly affected by damping if they propagate further than one pressure scaleheight $H$ over a damping time,
\Beq
\frac{\omega}{k_r H} \gtrsim K_{\rm rad} k_r^2,
\Eeq
where $K_{\rm rad}$ is the radiative diffusivity. This is a conservative estimate: waves need to propagate many scaleheights for them to reach the surface without damping. We can rewrite this in terms of the radiative luminosity using $L_{\rm rad}\sim (4\pi r^2) K_{\rm rad} \rho N^2 H$. Then a requirement for waves to not experience significant damping is
\Beq
\frac{4\pi r^2 \rho \omega^4}{Nk_h^3} \gtrsim L_{\rm rad}.
\Eeq
If waves are radiatively damped, they will not reach sufficient amplitudes to break. So combining this with equation~\ref{eqn:damped}, we find a requirement for waves to break is
\Beq\label{eqn:break}
L_{\rm wave} \gtrsim L_{\rm rad}.
\Eeq
Since the wave luminosity is significantly smaller than the convective luminosity, this relation is not satisfied in almost all stars \citep[some stars in the last year of their life may have $L_{\rm wave}>L_{\rm rad}$, see][]{Quataert2012}. Thus, traveling waves will not experience wave breaking. Note, however, that many numerical simulations artificially increase the convective luminosity by many orders of magnitude \citep[e.g.,][]{Rogers2013,Jones2017,Edelmann2019}, which likely explains reports of breaking waves generated by convection in \citet{Rogers2013} and \citet{Edelmann2019}. Artificially increasing the convective (and hence wave) luminosity makes it difficult to study wave excitation and propagation, because the wave physics can be very different.

\subsection{Nonlinear stability of standing waves}\label{sec:nonlinear damping}

Figure~\ref{fig:transfer} shows that waves excited near the star's g-mode frequencies can experience resonant amplification.
Because the resonant waves are coherent for many wave periods, weakly nonlinear effects can build up over time, and lead to significant transfer of energy between frequencies without necessitating wave breaking.

However, the observed luminosity fluctuations are very small. Even if the entire luminosity variation was due to standing modes, the modes would be \textit{stable to nonlinear instability} for most low-$\ell$ standing modes in $10 M_\odot$ ZAMS stars. As discussed in \citet{Weinberg2012}, nonlinear instabilities of a ``parent'' mode with frequency $f$ and amplitude $\xi_h k_h$ requires the existence of two other modes satisfying
\Beq\label{eqn:nonlinear instability}
\xi_h k_h f \kappa > \sqrt{\gamma_1 \gamma_2} \left[1 + \frac{\delta f^2}{(\gamma_1+\gamma_2)^2}\right]^{1/2},
\Eeq
where $\kappa$ is a spatial overlap integral, $\gamma_1$ and $\gamma_2$ are the damping rates of the two ``daughter'' modes, and $\delta f$ is the frequency difference between the parent and daughter modes, all measured in units of ${\rm d}^{-1}$. The difference between the $\ell$ of the daughter modes can be no greater than the $\ell$ of the parent mode.

\begin{figure}
  \centering
  \includegraphics[width=\linewidth]{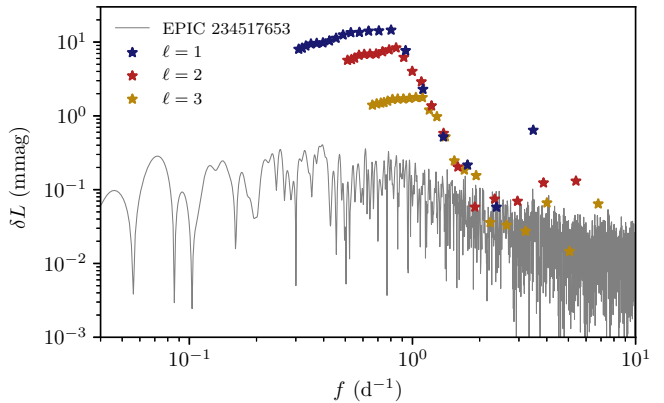}
  \caption{The stability threshold of weakly nonlinear standing waves in comparison to the observed spectrum for EPIC 234517653 from \B. The symbols show the luminosity variation of a marginally stable standing mode of a given $\ell$. Most modes require greater-than-observed luminosity variations to be unstable to weakly nonlinear instabilities.}
  \label{fig:break}
\end{figure}

We calculate the threshold for nonlinear instability by finding the smallest possible $\xi_h k_h$ satisfying equation~\ref{eqn:nonlinear instability} for each mode. We search over modes with $\ell\leq 20$ with daughter mode frequencies less than the parent mode frequency so they can efficiently damp. We assume that the spatial overlap integral $\kappa$ is small if the difference between the radial mode orders is greater than two; otherwise we take $\kappa=1$. In figure~\ref{fig:break}, we plot the surface luminosity fluctuations associated with these marginally stable modes. All but one $\ell=1$ mode would be nonlinearly stable, even if we assume the observed fluctuations are entirely due to those modes. We also find nonlinear stability for modes with $f \leq 1\, {\rm d}^{-1}$.

Less massive and more evolved stars have more weakly damped modes, so some standing modes experience nonlinear instabilities in these stars. We also note that some stars in \B\ show several peaks, likely due to linearly unstable g-modes. In many cases, those peaks are at amplitudes similar to the marginally stable amplitude calculated in Figure~\ref{fig:break}.  Many other stars show signatures of nonlinear wave interaction \citep[e.g.,][]{Degroote2009,Bowman2016}. In pulsating white dwarfs (ZZ Ceti stars), the mode amplitudes are also believed to be limited by weak non-linearities \citep[e.g.,][]{Wu2001,Luan2018}. However, in none of these cases is there any indication that these interactions can produce the ``red noise'' spectra observed in \B.

\section{Subsurface convection}

We have argued that the observed low-frequency variability is massive stars is unlikely due to internal gravity waves generated by core convection. Another possible source of the variability is surface phenomena.

Stars with spectral types later than F have well-developed surface convection zones driven by hydrogen ionization. In these stars, turbulent convection results in a specific photometric signature: granulation. In the power spectrum, granulation shows up as an excess of power at low frequencies (red noise).  3D radiative hydrodynamical simulations of surface convection can match observations of the disk-integrated intensity of main sequence stars and red giants \citep[e.g.][and references therein]{Samadi2013,Samadi2013b}. Convective surface velocity fluctuations also affect stellar spectra. If the surface velocity fluctuations are correlated over the entire line-forming region, this is referred to as macroturbulence. Otherwise it is referred to as microturbulence.  

Both micro- and macro- turbulence have also been observed in OB stars, with amplitudes ranging from a few to tens of ${\rm km}\,{\rm s}^{-1}$.  
These stars are not convective at the surface, but possess subsurface convective layers due to partial ionization of helium and iron group elements  \citep{Cantiello2009,Cantiello2019}. 
\citet{Cantiello2009} and \citet{Grassitelli2015,Grassitelli2015b} show that the presence and vigor of subsurface convection is correlated to spectroscopically-derived surface velocity fluctuations in OB stars. This is true for all stars except those with magnetic fields strong enough to substantially affect the iron subsurface convection zone \citep{Sundqvist2013}.
This suggests that subsurface convection could play an important role in inducing stellar surface variability, even for OB stars\footnote{At the same time, the collective effect of gravity modes could also explain the macroturbulent velocities in these stars \citep{aerts:2009}}.

It is thus natural to ask if the observed low-frequency variability in massive stars is also caused by subsurface convection.
This was discussed in \citet{Bowman2019AA}, which showed that the variability in massive stars has a characteristic frequency $\nu_{\rm char}$ of a few tens of $\mu$Hz, with little dependency on the star's position on the HR-diagram. This is inconsistent with the ``KB-scaling'' of convective frequency with stellar properties derived in \citet{Kjeldsen1995} (see Figure~8 in \B). 

However, \citet{Kjeldsen1995} derived this scaling for \textit{surface} convection.
The KB-scaling matches observations of stars with surface temperatures below approximatively $10-11$\,000 K, which have surface convection driven by  hydrogen and helium ionization.
For hotter stars, convection occurs below the surface around temperatures corresponding to partial ionization of HeII ($\approx$ 45kK) and Fe ($\approx$ 150kK).
These \textit{subsurface} convective regions (HeIICZ and FeCZ) occur at the specified temperatures; if a star expands, they will move to deeper mass coordinate.
As such, they do not directly trace stellar surface properties, and one does not expect their induced surface photometric fluctuations to follow the KB-scaling.
Instead, models predict a fairly constant value of their characteristic frequency as function of the stellar parameters, with typical frequencies $6-60$ $\mu$Hz \citep{Cantiello2009,Cantiello2011,Cantiello2019}, consistent with the range of characteristic frequencies observed in the \citet{Bowman2019AA} sample. Hence, 
we argue the distribution of characteristic frequencies of the observed low-frequency variability may support a subsurface origin.

The low-frequency variability is also observed in low-metallicity LMC stars (\B).
Since the occurrence of the FeCZ is metallicity dependent, this could rule out subsurface convection as the cause of the variability.
In 1D stellar evolution calculations, the FeCZ occurs above a luminosity of L$\approx10^{3.2}\Lsun$ for solar metallicity stars, corresponding to a zero age main sequence star of about 7$\Msun$.
At the lower metallicity of the LMC, the FeCZ appears at higher luminosities (L$\approx10^{3.9}\Lsun$), corresponding to a zero age main sequence star of about 11$\Msun$ \citep{Cantiello2009}. 
However, it is unclear if the objects in the TESS LMC sample of \B\ are hot, main-sequence stars probing this FeCZ transition luminosity. For example, if
the sample contains mostly luminous giants and supergiants, then the stars could be either above the FeCZ luminosity threshold for LMC metallicity, or posses surface convection due to H and He ionization. This would also be consistent with the much larger amplitude of the photometric variability of the LMC sample, as expected in more massive, brighter stars, independent of the scenario considered (subsurface convection or core IGWs). 

Finally, some of the stars showing low-frequency, stochastic variability are low-mass, main-sequence A stars. These are not expected to show the presence of an FeCZ. However, they are affected by other (sub)surface convection zones due to the ionization of H and He \citep{Cantiello2019}. While the amplitude of the velocity fluctuations induced by these convective regions tend to be smaller than for the FeCZ, characteristic frequencies are also in the range of tens of $\mu$Hz.

Therefore, subsurface convection remains a viable explanation for the observed variability. More work is required to test this hypothesis and firmly establish---or rule out---a connection between subsurface turbulent convective motions and the wide-spread, low-frequency photometric variability observed in massive stars. 

\section{Summary}

\begin{itemize}
\item We argue that the low-frequency variability presented in B19 is not due to linearly propagating waves from the core because:
\begin{itemize}
\item At low frequencies, linear waves have a smooth spectrum, but very little surface luminosity variation due to strong radiative damping. This is not observed.
\item Linearly propagating waves at frequencies $f\gtrsim 0.5 \, {\rm d}^{-1}$ form many standing modes (Figure~\ref{fig:spectra}), which are also not observed.
\end{itemize}
\item Wave propagation can be approximated as linear because:
\begin{itemize}
\item Low frequency traveling waves do not break.
\item If high frequency variability was due to standing modes, many would be at such low amplitude they would be stable to nonlinear interactions.
\end{itemize}
\item We conclude that the spectra presented in B19 are not due to internal gravity waves generated by core convection.
\item However, almost all massive stars have subsurface convection zones which could produce the observed low-frequency variability.
\end{itemize}

\acknowledgments

We are grateful for extremely useful discussions with Conny Aerts, Dominic Bowman, and Siemen Burssens regarding the observations presented in \B. We would like to thank Rich Townsend for valuable discussions about GYRE and stellar oscillation modes. We would also like to thank Tamara Rogers, Rathish Ratnasingam, and Philipp Edelmann for useful conversations about wave nonlinearity. We would also like to thank the anonymous referee who pointed out some additional reasons why the observations of \B\ are unlikely to be due to waves generated in the core. DL is supported by a PCTS and Lyman Spitzer Jr.~fellowship. DL would like to thank his fellow members of the 10$^{\rm th}$ New York County Grand Jury \#6 of 2019 for their amiable company whilst he composed this paper during his four weeks of jury service. The Center for Computational Astrophysics at the Flatiron Institute is supported by the Simons Foundation. This work has\ benefited from many conversations at meetings supported by\, Grant GBMF5076 from\, the\, Gordon and Betty Moore Foundation. EQ thanks the Princeton Astrophysical Sciences department and the theoretical astrophysics group and Moore Distinguished Scholar program at Caltech for their hospitality and support. LAC, BF and MLB acknowledge funding by the European Research Council under the European Union's Horizon 2020 research and innovation program through Grant No.681835-FLUDYCO-ERC-2015-CoG. This work was performed in part under contract with the Jet Propulsion Laboratory (JPL) funded by NASA through the Sagan Fellowship Program executed by the NASA Exoplanet Science Institute. BJSP acknowledges being on the traditional territory of the Lenape Nations and recognizes that Manhattan continues to be the home to many Algonkian peoples. We give blessings and thanks to the Lenape people and Lenape Nations in recognition that we are carrying out this work on their indigenous homelands.

\bibliographystyle{apj}
\bibliography{waves}

\end{document}